# Study of the feasibility of a compact gamma camera for real-time cancer assessment

**L. Caballero**

*Instituto de Física Corpuscular - CSIC - University of Valencia; C/Catedrático José Beltrán, 2; E-46980; Paterna; Spain*
*E-mail*: Luis.Caballero@ific.uv.es

ABSTRACT: Results from the simulations of a Compton gamma camera based on compact configuration of detectors consisting in two detection modules, each of them having two stages of high-resolution position- and energy-sensitive radiation detectors operated in time-coincidence are presented. Monolithic scintillation crystals instead of pixelated crystals in order to reduce dead areas have been simulated. In order to study the system feasibility to produce real-time images, different setups are considered. Performance in terms of acquisition times have been calculated to determine the real-time capabilities and limitations of such a system.

# Contents

## 1. Introduction and motivation

All the devices involved in molecular breast imaging are based on detection of gamma-rays emitted by a radiopharmaceutical substance injected in the patient for which the tumour is avid or which drainage reproduces the cancer dissemination path. A device capable of obtaining metabolic information in real-time and use it to guide the biopsy procedure will be a change of paradigm in breast cancer assessment. Targeted biopsies based on the tumour heterogeneity shown by a gamma imaging device will open the door to personalized diagnosis and treatments, thus improving the patient prognosis.

Gamma cameras play a central role in molecular imaging, being capable of depicting the radioisotopes distribution in the patient previously injected. From the initial Anger-based conventional gamma cameras to the dedicated breast PEMs there have been different technical approaches in order to obtain information about the radioisotope distribution in the patient. The large distances to the breast in the case of the conventional gamma cameras was solved by the development of small compact gamma cameras [1] based on the use of large arrays of pixelated scintillation crystals (CsI or NaI) coupled to position-sensitive photomultipliers (PSPMTs). Despite this technology allowed a closer positioning of the detection system to the breast, they are too bulky to permit any access in case of a biopsy procedure is desired.

Positron Emission Tomography (PET) scans are recommended for cancer staging because of their capacity to determine whether the cancer has spread or not. Dedicated breast PETs, also known as PEMs, have demonstrated better performance than whole-body-PETs in terms of spatial resolution, being able to depict tumours of less than 1 cm, which becomes especially important to assess breast cancer at early stages. There has been attempts to adapt the biopsy procedure to the PEM imaging technique, such as Solo II PEM [2], but none of the state-of-the-art devices is capable of co-relate the images with morphologic information in real-time, so a realistic needle guiding is not possible.

In this report, I study the feasibility of using a Compton gamma camera to generate images in real-time of the radiotracer injected into the patient. Such a device will permit its coupling with a, for instance, ultrasound device in order to co-register the images and guide the biopsy needle to the lesion region which presents a higher FDG uptake. I also determine the minimum time required to accumulate enough statistics to reconstruct images with enough quality to depict the tumours radiotracer uptake.

## 2. Tumour simulation

The total amount of FDG accumulation in tumours depends on the tumour metabolism. PET scanners are designed to measure the in vivo radioactivity concentration [kBq/ml], which is directly linked to the FDG concentration, being the relative tissue uptake of FDG of interest. Typical tumour uptake values range from 5 to 10 kBq/cc as can be seen in Fig.1 (source: [3]). As it can be observed, 10 min after the injection time, the uptake is kept constant in the tumour whereas the tissue shows a reduction over time.

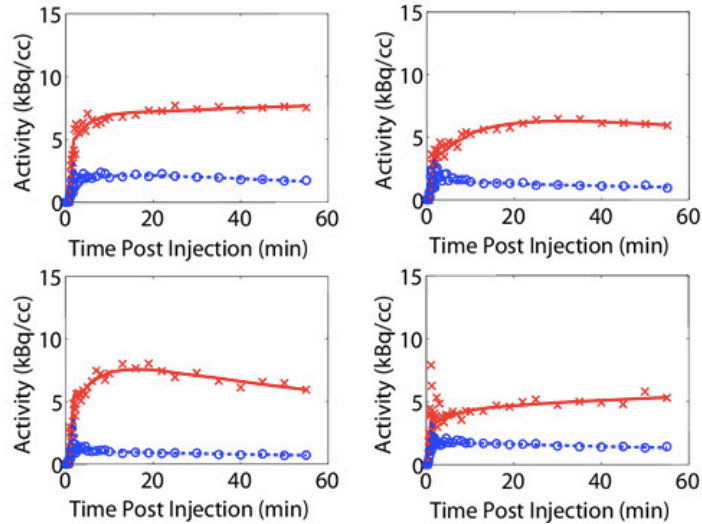

**Fig.1: Dynamic measured and modelled tumour uptake values in PET/CT for four representative patients (source: [3]).**

## 3. Compton gamma camera design and Monte Carlo simulation

The gamma detection system simulated is a Compton gamma camera based on compact configuration of detectors consisting in two detection modules, each of them having two stages of high-resolution Position- and energy-Sensitive radiation Detectors (PSDs) operated in time-coincidence. Each PSD has a monolithic scintillation crystal and a SiPMTs in order to allow them to be placed at short distances. The closer PSDs to the radiactive source are called the *scatterers*, and the rest are called the *absorbers*. In this solution, monolithic scintillation crystals are used instead of pixelated in order to maximize the detection sensitivity. The scintillation crystals simulated were $LaBr_3(Ce)$ of $20\times60\times10$ mm$^3$ for the *scatterers* and $20\times60\times20$ mm$^3$ for the *absorbers*. Each PSDs scatterer and absorber were separated 20 mm. Monte Carlo simulation was performed using the Geant4 code based program [4]. In this simulation, no optical physics was considered for the scintillation light produced by the gamma-rays impinging in the crystal. Position and energy deposited were extracted for each gamma-ray interaction within the scintillation crystals, so no image compression due to the finite size of the crystals is considered in the present approach. A basic geometric based backprojection algorithm was used to reconstruct the images, in which the hyperbola resulting from the intersection of the Compton cone with the ultrasound field of view is calculated (see Fig 2). As I show in Fig3, a punctual isotropic source emitting gamma-rays of 511 keV simulating a tumour with an uptake of 5 kBq of $^{18}$F-FDG has been simulated. The source was placed in vacuum and no surrounding tissue assumed. Tumour depth distance to the gamma camera was of 3 cm.

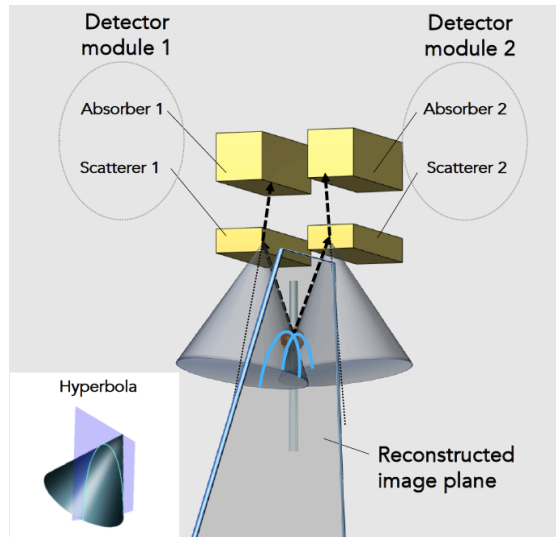

**Fig.2: Reconstruction image plane. Intersection of the Compton cone with the reconstruction plane is defined by a hyperbola. Accumulation of hyperbolas generates the gamma image depicting the gamma-ray source.**

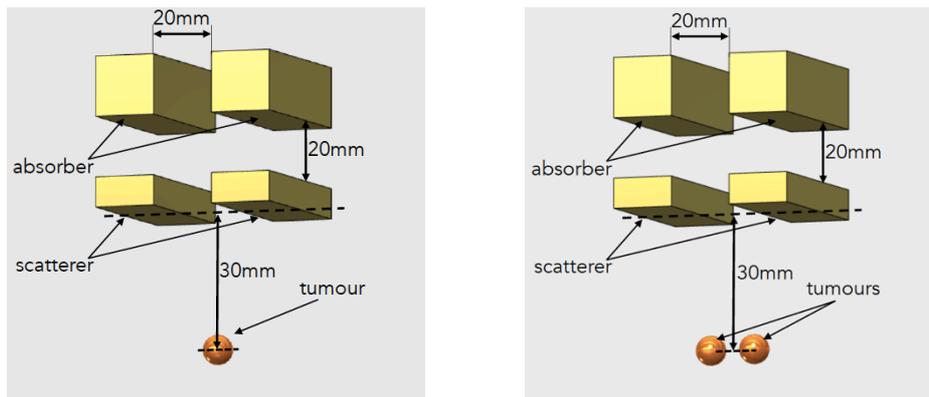

**Fig.3: Schematics of the simulated setup for one (left) and two punctual tumours (right).**

The two setups simulated in this work are shown in Fig.3. First, and for simplicity, a punctual source emitting gamma-rays of 5 kBq positioned in the plane in between the detector modules was simulated. Tumour depth in the patient was assumed to be 1.5 cm respect to the skin, which was assumed as well at a distance of 1.5 cm respect to the scatterer frontal-face plane, being the total distance 3 cm from the tumour to the detectors plane. In Fig.4 I show the reconstructed images depending on the acquisition time. As can be seen, only ~5 seconds are required to generate an image with enough resolving power to clearly identify the tumour in the gamma image. Despite the more statistics, the better the tumour is defined, there is no significant change in the tumour definition beyond 15s of acquisition.

In order to evaluate the resolving power, two punctual sources separated by 10 mm and emitting gamma-rays of 511 keV simulating two tumours with an uptake of 5 kBq/cc of $^{18}$F-FDG each have been simulated. Tumours depth distance to the gamma camera was also 3 cm as previously explained (see Fig3).

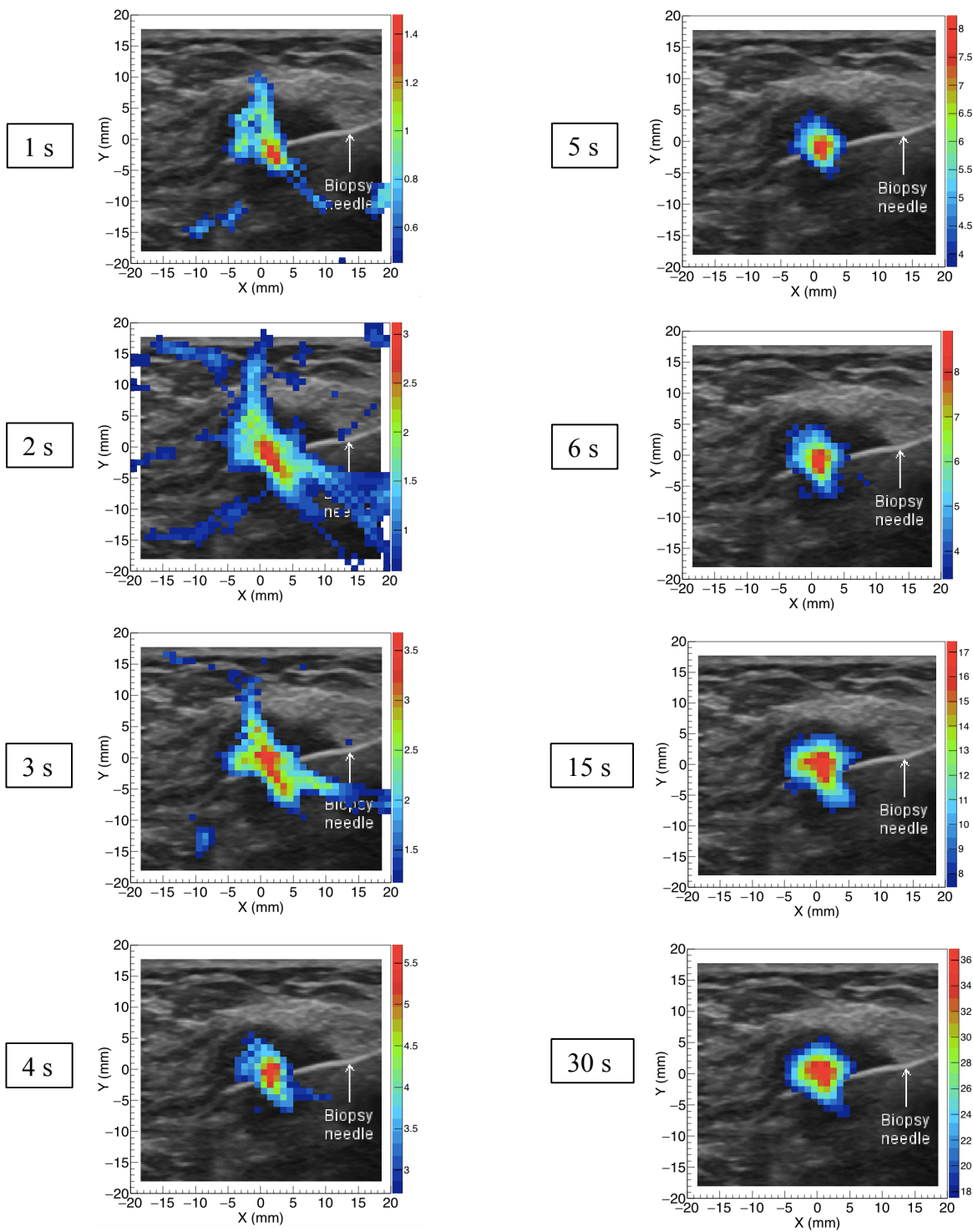

**Fig. 4**: Evolution of the gamma images reconstructed for a simulated tumour of 5 kBq/cc for different durations of the scan (background ultrasound image only used as a reference).

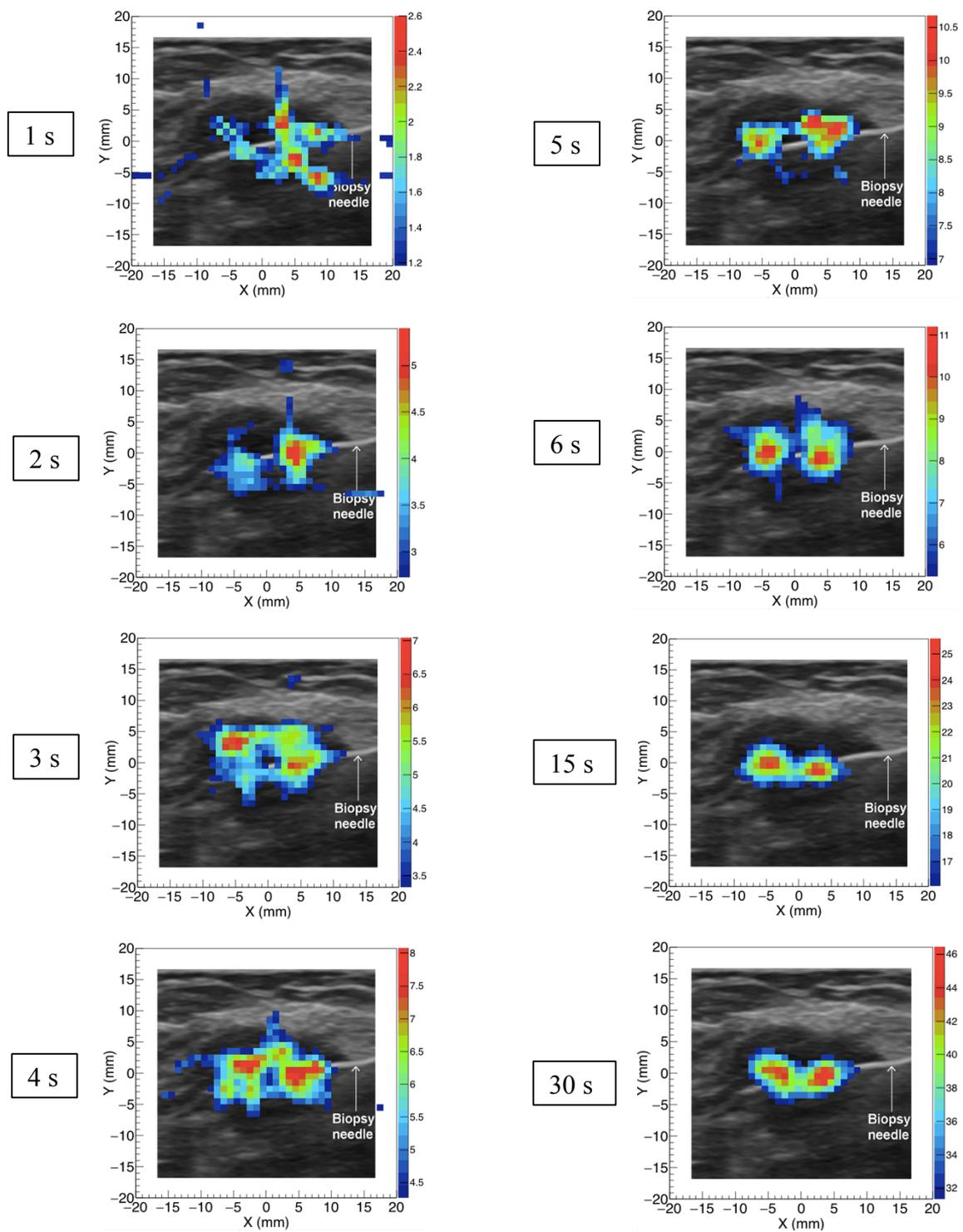

Fig. 5: Evolution of the gamma images reconstructed for two simulated tumours of 5 kBq/cc each for different durations of the scan (background ultrasound image only used as a reference).

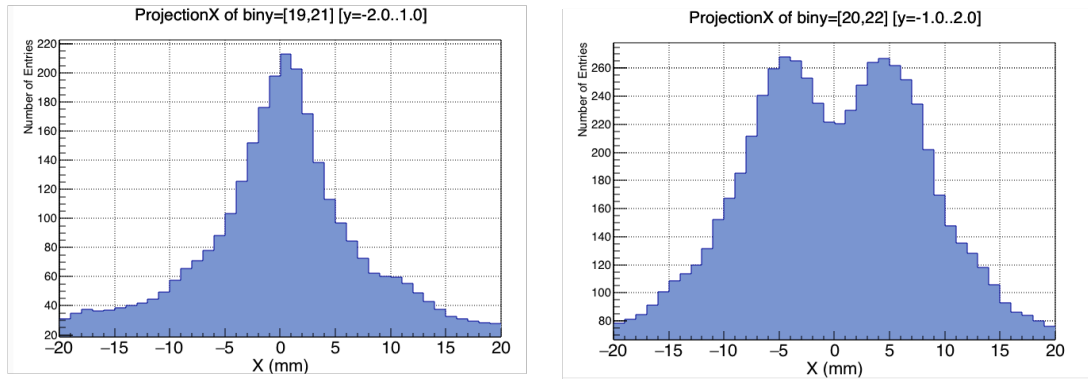

**Fig. 6: Projections of the reconstructed images (acquisition of 60 s) for one punctual source (left) and two punctual sources (right).**

In Fig.6 I plot the images projection two show the spatial resolution of the system, being 9.5 mm at FWHM which, as also shown is enough to resolve the two tumours case.

## 4. Conclusions

A very schematic simulation has been performed to evaluate the resolving power of a Compton gamma camera based on a particular configuration that permits to embed an ultrasound transducer in between the detector modules. Despite the rough approach both in the optical physics and tumour shape and activity distribution simulated, a good estimation of the feasibility of such a gamma detector design in terms of reconstructing gamma images in real-time is obtained. Also, further improvement will be obtained when simulating different scintillation crystals. Implementation of more elaborated and powerful image reconstruction algorithms such as MLEM, OSEM, etc. or machine learning will produce much better results even for shorter scan times. Thus, a gamma detection system such as the studied is feasible of producing gamma images of tumours in a realistic real-time mode making feasible the biopsy needle guidance.

## References

[1] Oncovision, GEM Imaging S.A, Spain. http://oncovision.com/sentinella/

[2] CMR Naviscan Corporation, California, USA. http://www.cmr-naviscan.com/naviscan-solo-2/

[3] Wangerin KA, Muzi M, Peterson LM, et al. Effect of $^{18}$F-FDG uptake time on lesion detectability in PET imaging of early stage breast cancer. *Tomography: a journal for imaging research*. 2015;1(1):53-60. doi:10.18383/j.tom.2015.00151.

[4] GEANT4 Collaboration. *GEANT4-a simulation toolkit.* Nuclear Instruments and Methods in Physics Research A, 506:250–303, July 2003.